\pdfoutput=1
\documentclass[11pt, a4paper, nonumbering, trilogo]{google}

\usepackage[numbers, square, comma, sort&compress]{natbib}
\usepackage{graphicx}
\usepackage{booktabs}
\usepackage{tabularx}
\usepackage{longtable}
\usepackage{array}
\uselogo{true}
\title{Agent-guided epitope excision enables economical \textit{de novo} binder design}
\correspondingauthor{junki@google.com, apawlosky@google.com}
\reportnumber{}

\renewcommand\today{2026-10-23}

\author[1,$\dagger$]{Jun W. Kim}
\author[2]{Artiom Myaskovsky}
\author[1]{Oscar Donosa}
\author[3]{Lawrence Chan}
\author[3]{Jamie Rogers}
\author[1]{Avinatan Hassidim}
\author[1]{Yossi Matias}
\author[1,$\dagger$]{Annalisa Pawlosky}

\affil[1]{Google Research}
\affil[2]{Google DeepMind}
\affil[3]{Google Cloud}
\affil[$\dagger$]{Corresponding author}

\begin{abstract}
\textit{De novo} protein design campaigns remain bottlenecked by the inference compute time required to generate and rank candidates, limiting wider functional exploration. Here, we introduce and evaluate an epitope excision approach that truncates targets to local interface windows to accelerate both backbone diffusion and structure prediction. Across 15 targets and 3,666 designs, excised windows closely reproduced full-target binding interfaces (median $\mathrm{C}\alpha$ RMSD of $0.37 \,\text{\AA}$) and made it practical to design binders against the 719-residue extracellular domain of fibroblast activation protein alpha (FAP) by cutting RFdiffusion and AlphaFold3 GPU time by $\sim  88\%$ in one of the epitopes. Pairing this workflow with an agentic optimization loop allowed the system to autonomously identify catalytic-site targeting strategies. Under a fixed compute budget, a human-in-the-loop (HITL) workflow raised the computational hit rate from 5.06\% to 9.33\% while further cutting GPU hours by $\sim  42\%$. We also found that omitting target multiple sequence alignments during refolding roughly substantially increased the apparent yield. Wet-lab validation with yeast display identified nanomolar binders from the generated designs with the HITL and the fully autonomous agent (human-out-of-the-loop, HOOTL) arms recovering more enriched designs than the PI-alone baseline. Together, target excision and agent-guided search offer a practical way to scale \textit{de novo} binder design to large, therapeutically relevant proteins.
\end{abstract}

\begin{document}
\maketitle

\section*{Introduction}

Structure-based \textit{de novo} binder design routinely generates thousands of candidate backbones and sequences per target, combining backbone generators such as RFdiffusion \cite{ref1} with sequence design tools like ProteinMPNN \cite{ref2} in end-to-end pipelines such as BindCraft \cite{ref3}. Because only a small fraction of generated candidates actually bind \textit{in vitro}, campaigns depend on an \textit{in silico} filtering step in which each candidate is refolded with its target (for example with AlphaFold3 \cite{ref4}, building on AlphaFold2 \cite{ref5}) and ranked by an interface-confidence metric. One such metric, ipSAE \cite{ref6}, is derived from AlphaFold predicted aligned error and has been shown to outperform earlier scores including pDockQ \cite{ref7} and pDockQ2 \cite{ref8}. How well this computational screen separates true binders from non-binders has been shown to correlate with the hit rate of downstream wet-lab testing.

The refolding step is also the primary compute bottleneck. AlphaFold3 has been demonstrated to be a reliable refolding method, and its runtime is governed by discrete token buckets, with complex lengths rounded up to the next multiple of 256 tokens. Many biologically important targets, including cell-surface receptors, multi-domain ectodomains, viral glycoproteins, and enzymes are much larger than the designed binder and the local patch it contacts. Consequently, most of the tokens processed during refolding belong to target regions far from the binding site. For a 719-residue enzyme ectodomain of a disease target, fibroblast activation protein $\alpha $ (FAP), repeated full-structure diffusion and full-complex refolding is prohibitively slow, even though the targetable epitope is a relatively small rim around the catalytic opening \cite{ref9}.

To get around the compute bottleneck, previous studies have targeted structurally independent folding units. However, many drug targets containing functional pockets are not confined to a single, small domain. Recent work has demonstrated that the global fold of a target may not be strictly necessary for generative design, as discontinuous structural fragments can successfully guide backbone generation \cite{ref10}. However, target truncation requires rigorously auditing of its downstream consequences for larger wet-lab campaigns. It remains unknown how aggressive cropping impacts the precise interface geometry and the ranking signal required to separate true binders from non-binders during \textit{in silico} triage. Furthermore, discontinuous 3D patches do not map cleanly to the discrete token-bucket economics of modern predictive triages like AlphaFold3, nor do they readily support the multiple sequence alignments (MSAs) often required for reliable refolding. This asymmetry motivates epitope excision: cropping the target down to the residues immediately surrounding the designed interface (the contact shell plus a tunable surrounding halo) and folding this local pocket in place of the intact protein. For excision to be useful in practice, it must preserve the local geometry of the designed interface, maintain the relative ranking of true binders against non-binders, and cross a token-bucket boundary to deliver actual wall-clock speedups. Instead of treating aggressive cropping as a free speedup, the first half of this study tests all three requirements systematically across a 15-target benchmark.

Beyond reducing fold times, excision turns otherwise intractable targets into manageable sub-problems. Running a productive campaign still requires iterative choices about where to target, how much of the surrounding structure to retain, and how to allocate compute across candidate backbones and sequences. Large language model (LLM) agents are well suited to this type of iterative, data-driven optimization. Equipped with external tools, state memory, and quantitative feedback, modern LLMs can plan multi-step workflows, launch jobs, inspect intermediate results, and adjust their search policy from round to round, closely mirroring how a researcher runs a design campaign. Similar multi-agent frameworks built on Gemini, such as Google's AI co-scientist, have recently been used to generate, critique, and refine experimentally validated biomedical hypotheses \cite{ref11}. \textit{De novo} binder design provides a clean testbed for evaluating how much autonomy to give such an agent, because campaign performance can be measured directly as the computational hit rate under a fixed structural threshold and a fixed GPU budget.

In the second half of this work, we hold the excision pipeline, scoring criteria, and total GPU budget constant, and vary the level of agent autonomy on FAP across three regimes: PI alone, a human--agent hybrid (Human-in-the-Loop, HITL), and a fully autonomous agent (Human-out-of-the-Loop, HOOTL). The two parts of the study address complementary questions: the 15-target benchmark defines the physical and statistical operating limits of epitope excision, and the FAP campaign tests how best to direct that pipeline in practice. We find that the human--agent hybrid produces the most high-confidence designs per GPU-hour. Along the way, we also observe that omitting target multiple sequence alignments (MSAs) substantially shifts AF3 interface-confidence scores on several targets, indicating that single-sequence refolding can be an effective option depending on alignment depth. Finally, we test these computational predictions experimentally through yeast display and identify nanomolar binders from the generated designs.

\section*{Results}

The output of a \textit{de novo} binder campaign depends heavily on computational triage. Because true binders are rare among generated candidates, every design must be refolded with its target and ranked by an interface-confidence metric before selecting sequences for synthesis. A natural way to try to improve this ranking is to spend more compute per candidate by folding each complex with multiple independent structure predictors and combining their scores. To test the upper bound of this approach, we built a multi-engine consensus model and evaluated it using leave-one-target-out cross-validation across the 15-target benchmark used throughout this study (3,760 designs from \cite{ref12,ref13}). In this setup, three distinct structure-prediction models (AlphaFold3, Boltz-2, and ESMFold2) supply interface-confidence metrics alongside biophysical features (PRODIGY interface $\Delta  G$, FreeSASA buried surface area, and EvoEF2 energy terms) to a stacked ensemble of linear, multilayer perceptron (MLP), and gradient-boosted models (\textbf{Supporting Fig. 1a}, \textbf{Supporting Table 1}). The resulting gain in discrimination is modest: median average precision (AP) increases from $0.54$ using raw AF3 ipSAE alone to $0.60$ for the full three-engine ensemble, and the two steps that require running additional folding models (Boltz-2 and ESMFold2) together contribute only $+0.02$ AP while roughly doubling GPU time per design (\textbf{Supporting Fig. 1b}). Adding more folding engines per candidate quickly hits diminishing returns.

If prediction accuracy per fold is difficult to improve by brute force, the more practical way to increase success rate is to increase screening throughput to enhance selection by reducing the cost of each fold without degrading binder ranking. Epitope excision (\textbf{Fig. 1}) approaches this by folding each candidate binder against a local crop of the target around the intended interface rather than the entire target protein. In the sections below, we test whether cropped targets preserve interface geometry and binder ranking across the 15-target benchmark \cite{ref12}, define the minimum crop size needed for reliable folding, and then apply the resulting pipeline to FAP.

\begin{figure}[!htbp]
\centering
\includegraphics[width=\textwidth]{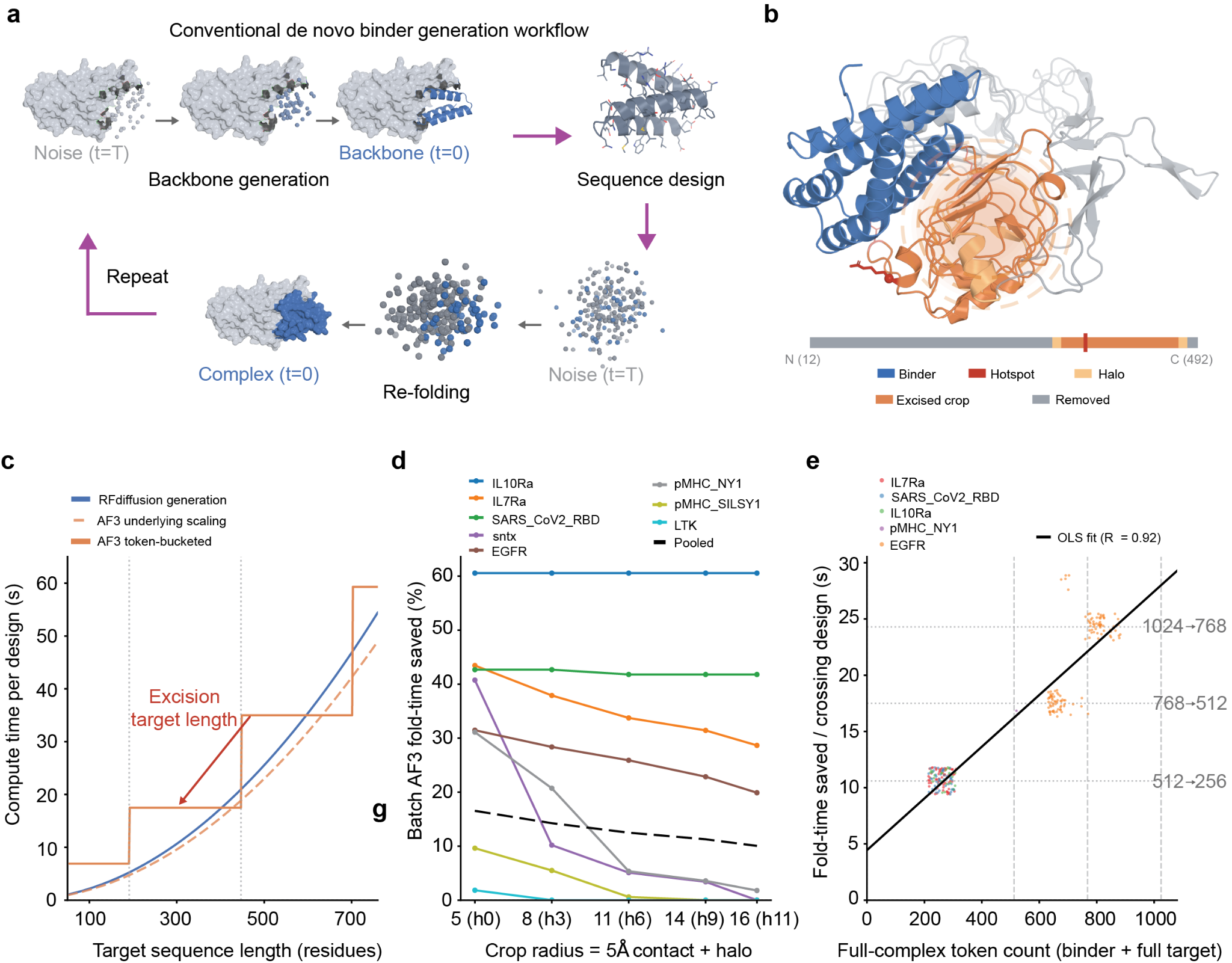}
\caption*{\textbf{Figure 1} | \textbf{Epitope excision and its compute economics.} \textbf{(a)} Conventional \textit{de novo} binder design: backbone generation (RFdiffusion), sequence design (ProteinMPNN), refolding of the binder against the \textit{full} target (AlphaFold3), evaluation and ranking (pLDDT, ipSAE, RMSD), iterated. \textbf{(b)} Epitope excision illustrated on an EGFR ectodomain complex (binder in blue). The binder contacts a localized surface patch mapping to a contiguous sequence window around hotspot Arg353 (red); excision retains that window plus surrounding halo residues (orange and light orange, $\sim  168$ residues total, with shading indicating optional halo width) and removes the remainder of the protein (grey) to leave a compact crop. \textbf{(c)} Compute time per design vs target length. Generation and folding scale quadratically ($\propto  L^2$); RFdiffusion is a continuous quadratic (blue), whereas measured AF3 refolding (orange) pays in discrete token buckets (rounded up to the next multiple of 256; $6.9/17.5/35.0/59.3 \,\text{s}$ for $256/512/768/1024$ tokens), a step around its quadratic envelope (dashed). Excision shortens the target dropping a design into a cheaper bucket. \textbf{(d)} Batch AF3 fold-time saved (\%) vs crop radius ($5 \,\text{\AA}$ contact + halo; halos $0/3/6/9/11 \,\text{\AA}$), one curve per saving target plus the pooled all-15 line (dashed). The bare pocket (halo 0) saves most; savings shrink as halo grows, and 7 of 15 targets are already in the smallest 256-token bucket. \textbf{(e)} For designs whose excised pocket crosses an AF3 bucket boundary at halo 11 ($n = 253$, 5 targets), fold-time saved per design vs full-complex token count (Pearson $r = +0.96$). Savings fall on discrete bands set by the boundary cleared ($512 \rightarrow  256$, $768 \rightarrow  512$, $1024 \rightarrow  768$, and a $\sim  28 \,\text{s}$ two-bucket band for the largest EGFR complexes). Savings from bucket-crossing scale with complex size.}
\end{figure}

In a standard binder campaign, the bulk of the compute budget goes to refolding. For instance, after generating backbones with RFdiffusion and assigning sequences with ProteinMPNN, a large number of candidates must be refolded with the target \cite{ref14} and evaluated by metrics, such as pLDDT, ipSAE, and interface RMSD (\textbf{Fig. 1a}). Refolding cost scales steeply with target length, especially for large multi-domain antigens, even though the binder only contacts a small surface patch. Epitope excision removes the non-contacting regions by retaining only the contiguous sequence window spanning the binder's contact residues plus a surrounding radial halo, and refolding the binder against this smaller crop (\textbf{Fig. 1b} shows one EGFR complex as an example, where the target is reduced to a 168-residue crop around hotspot Arg353; EGFR crops vary widely with the targeted epitope, with a median of 374 residues at the smallest halo). Because AF3 pads inputs to discrete token buckets (multiples of 256), trimming the target can drop a complex into a smaller bucket and substantially reduce wall-clock runtime (\textbf{Fig. 1c}).

Whether cropping actually reduces runtime depends on the initial size of the target and the radius of the retained halo (\textbf{Fig. 1d,e}). Across the 15-target benchmark, pooled batch fold-time savings are highest for minimal pockets (halo 0) and decrease as the halo widens, while 7 of the 15 targets show no speedup because their full-length complexes already fit within the smallest 256-token bucket (\textbf{Fig. 1d}). Among the designs that do drop into a smaller bucket ($n = 253$ across 5 targets), per-design time savings correlate strongly with full-complex token count (Pearson $r = +0.96$), grouping into discrete bands determined by which bucket boundary is crossed (\textbf{Fig. 1e}). The fraction of designs that cross a bucket boundary decreases monotonically with crop radius (\textbf{Supporting Fig. 2a}), and the wall-clock time saved per crossing is largest for big antigens like EGFR (\textbf{Supporting Fig. 2b}).

\begin{figure}[!htbp]
\centering
\includegraphics[width=\textwidth]{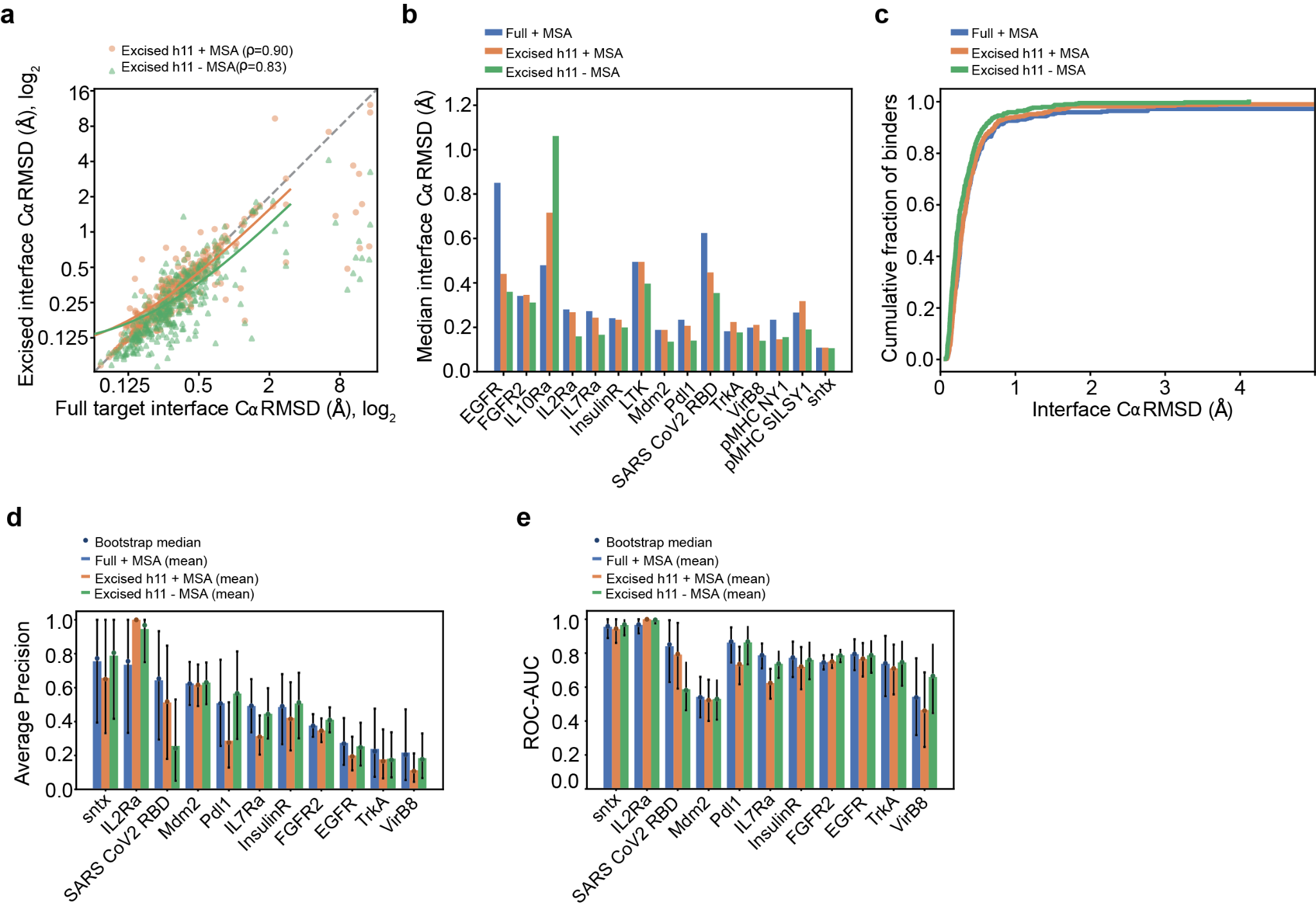}
\caption*{\textbf{Figure 2} | \textbf{Excision preserves designed interface geometry and ranking across 15 targets.} \textbf{(a)} Excised-pocket versus full-target interface-core $\mathrm{C}\alpha$ RMSD across experimentally confirmed binders for both excised conditions (halo-11 +MSA in orange; single-sequence no-MSA in green triangles). The diagonal indicates exact agreement with the intact target, and solid lines show ordinary least-squares fits within the valid $< 3 \,\text{\AA}$ regime. Agreement with full-target folds: +MSA Spearman $\rho  = 0.91$, $R^2(< 3 \,\text{\AA}) = 0.74$; no-MSA $\rho  = 0.83$, $R^2 = 0.54$. \textbf{(b)} Per-target median interface RMSD comparing full-target, excised+MSA, and excised no-MSA folds. \textbf{(c)} Empirical cumulative distribution function (ECDF) of interface RMSD across all three conditions. \textbf{(d)} Per-target average precision (AP) and \textbf{(e)} ROC-AUC of $\mathrm{ipSAE}_{\max}$ comparing full-target+MSA (blue), excised halo-11+MSA (orange), and excised no-MSA (green) across all designs on the informative target subset (error bars show 95\% design-resampling bootstrap intervals). The full-target bar for VirB8 (asterisk) shows our reproduced AF3 refold of all 99 designs (AP $0.18$, AUC $0.54$ across seeds).}
\end{figure}

For these compute savings to be useful, the cropped target must fold into the same binding interface as the full-length protein. We evaluated geometric fidelity by measuring the target interface-core $\mathrm{C}\alpha$ RMSD between the refolded complex and the original design model, using a $3 \,\text{\AA}$ threshold as a validity gate. Across the full excised benchmark ($11 \,\text{\AA}$ halo, 15 targets, 3,666 designs), the pooled median interface RMSD is $0.37 \,\text{\AA}$ (mean $0.53 \,\text{\AA}$), with 88.2\% of designs falling below $1 \,\text{\AA}$ and 93.7\% below $2 \,\text{\AA}$; per-target medians range from $0.20 \,\text{\AA}$ (Mdm2 and LTK) to $0.71 \,\text{\AA}$ (EGFR), well below the $3 \,\text{\AA}$ threshold. Because the source benchmark cached only scalar interface scores for full-target folds rather than full atomic coordinates, running a direct design-by-design RMSD comparison (\textbf{Fig. 2a--c}) required refolding the 394 experimentally confirmed binders against their intact targets (393 had a measurable interface core). On proven binders, excised crops match the intact targets closely: median interface RMSD is $0.29 \,\text{\AA}$ for excised targets versus $0.29 \,\text{\AA}$ for full targets, with 98.5\% and 97.2\% of binders passing the $3 \,\text{\AA}$ gate, respectively. Refolding the same excised pockets in single-sequence mode (no MSA) yields nearly identical geometry, with a median RMSD of $0.23 \,\text{\AA}$, 99.5\% of binders under $3 \,\text{\AA}$, Spearman $\rho  = 0.83$ against full-target RMSD, and only 4 of 393 gate calls differing from the +MSA excised condition (\textbf{Fig. 2a--c}). Per-binder RMSD agreement between excised (+MSA) and full-target folds is high (Spearman $\rho  = 0.91$, $n = 393$; within the valid $< 3 \,\text{\AA}$ subset, $n = 381$, Pearson $R^2 = 0.74$), with 98.2\% agreement on the $3 \,\text{\AA}$ gate (\textbf{Fig. 2a}). The few high-RMSD outliers, mostly on EGFR, fail in both full-target and excised refolds.

We next tested whether target excision affects the ability of $\mathrm{ipSAE}_{\max}$ to rank true binders above non-binders, treating each target as the unit of replication across all designs ($n_\mathrm{pos} \ge  5$, giving 11 informative targets). Comparing full-target+MSA folds against excised halo-11+MSA folds, mean AP goes from $0.47$ (full) to $0.40$ (excised), for a mean $\Delta $AP of $-0.065$ (paired-bootstrap 95\% CI $[-0.130, +0.019]$), while mean ROC-AUC goes from $0.78$ to $0.73$, for a mean $\Delta $ROC-AUC of $-0.048$ (95\% CI $[-0.083, -0.017]$). This drop is not uniform across the benchmark; it is driven primarily by a few targets with shallow MSAs or sensitive local geometry, specifically Pdl1 ($0.48$ to $0.25$) and IL7Ra ($0.48$ to $0.30$), whereas most other targets remain within their bootstrap confidence intervals and IL2Ra improves slightly (\textbf{Fig. 2d,e}).

Refolding the same excised pockets in single-sequence mode (no MSA) indicates that ranking drop on cropped targets can be influenced by the MSA. Removing the target MSA raises mean excised AP from $0.401$ to $0.454$, bringing it back within $0.012$ of the full-target+MSA baseline ($\mathrm{mean } \Delta (\mathrm{no-MSA} - \mathrm{excised+MSA}) = +0.053 \,\text{AP}$, paired-bootstrap 95\% CI $[-0.026, +0.126]$; $+0.036$ ROC-AUC; \textbf{Fig. 2d,e}, green). This recovery may depend on alignment depth: dropping the MSA improves ranking on targets with deep alignments (Pdl1 rises from $0.25$ to $0.54$, IL7Ra from $0.30$ to $0.43$, sntx from $0.60$ to $0.76$, and InsulinR from $0.39$ to $0.50$), likely because a deep target MSA can inflate baseline interface confidence for decoys when the pocket fold is simple. Conversely, removing the MSA cuts ranking performance in half on SARS\_CoV2\_RBD ($0.49$ to $0.24$), a shallow-MSA target that relies on coevolutionary constraints to fold its interface properly. Whether to include a target MSA during excised refolding should therefore be chosen on a per-target basis (because ranking metrics require both positive and negative examples, \textbf{Fig. 2d,e} includes all designs rather than only confirmed binders). These trends hold when complexes are scored using $\mathrm{ipSAE}_{\min}$, the more conservative chain-pair direction used in the source benchmark: absolute AP values are slightly higher across all three conditions (mean AP $0.510$ full, $0.451$ excised+MSA, and $0.497$ no-MSA), while the relative differences remain nearly unchanged ($\Delta (\mathrm{excised+MSA} - \mathrm{full}) = -0.058 \,\text{AP}$ and $-0.047$ ROC-AUC; $\Delta (\mathrm{no-MSA} - \mathrm{excised+MSA}) = +0.045 \,\text{AP}$ and $+0.039$ ROC-AUC; \textbf{Supporting Fig. 3}).

\begin{figure}[!htbp]
\centering
\includegraphics[width=\textwidth]{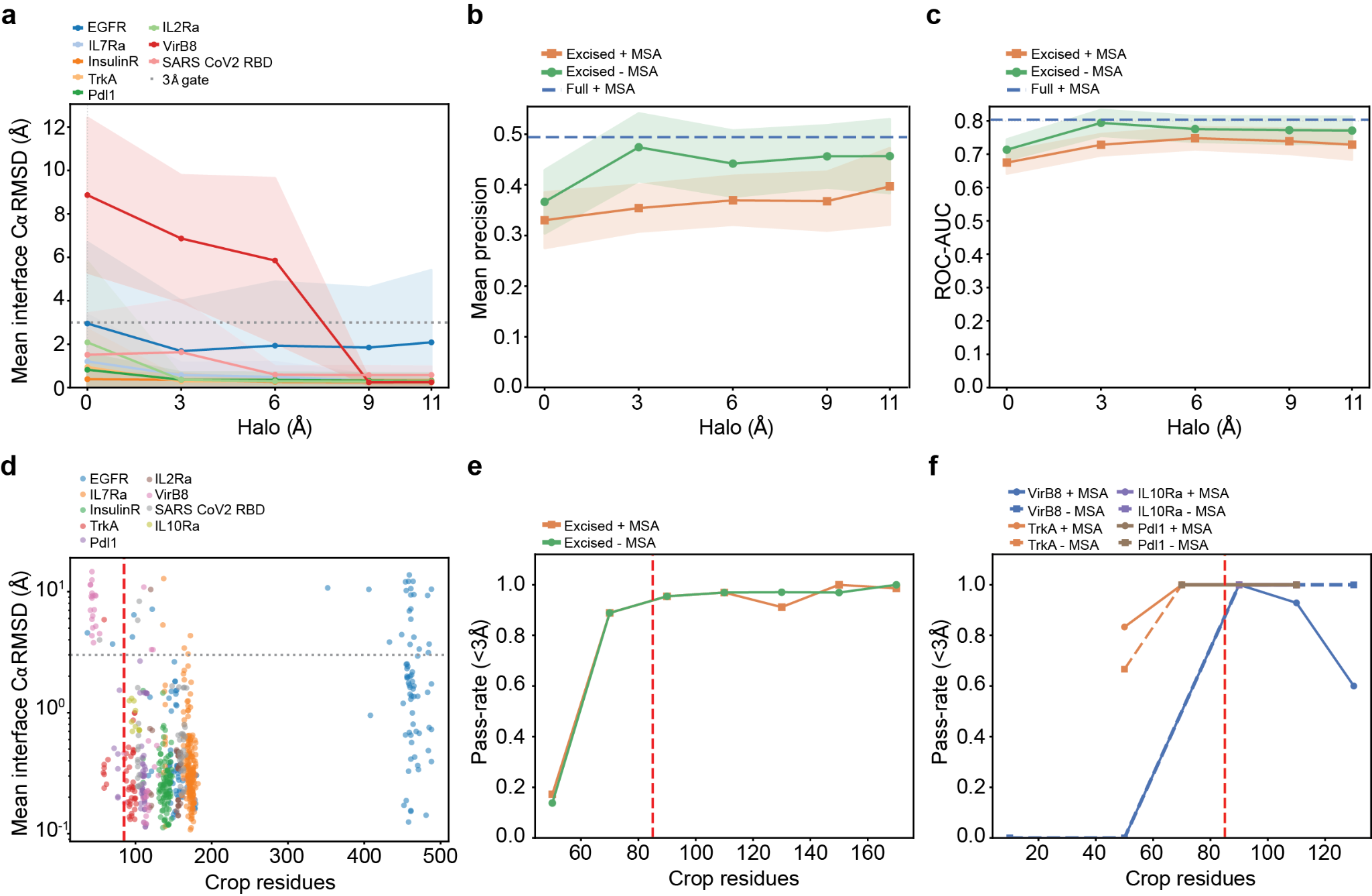}
\caption*{\textbf{Figure 3} | \textbf{Interface fidelity, ranking, and the crop-size floor define a safe operating window for excision.} \textbf{(a)} Interface $\mathrm{C}\alpha$ RMSD across halo radii ($0, 3, 6, 9,$ and $11 \,\text{\AA}$) for experimentally confirmed binders, plotted per target ($\mathrm{mean} \pm  1 \,\text{SD}$; dotted line indicates the $3 \,\text{\AA}$ validity gate). \textbf{(b,c)} Ranking accuracy across crop radii (halos of $0, 3, 6, 9,$ and $11 \,\text{\AA}$, corresponding to total radii of $5, 8, 11, 14,$ and $16 \,\text{\AA}$) scored on $\mathrm{ipSAE}_{\max}$ and matched design-for-design across the 11 well-powered targets (number of binders $\ge  5$; shaded bands show $\pm  1 \,\text{SE}$ across targets): mean AP \textbf{(b)} and ROC-AUC \textbf{(c)} for excised+MSA (orange) versus excised no-MSA (green), compared against the full-target+MSA baseline (dashed; AP $0.49$, ROC-AUC $0.80$) computed on the halo-matched design subset; $0.47/0.78$ over all designs in \textbf{Fig. 2}. \textbf{(d--f)} Absolute crop-size floor for confirmed binders as a function of retained target residue count (dashed red line marks the 85-residue threshold). \textbf{(d)} Per-fold interface $\mathrm{C}\alpha$ RMSD (log scale) versus retained residue count (+MSA), colored by target; crops shorter than 85 residues fall into the $> 3 \,\text{\AA}$ failure zone (shaded red, primarily VirB8), while larger crops fold reliably. \textbf{(e)} Confirmed-binder pass rate ($< 3 \,\text{\AA}$) across crop-size bins for excised+MSA (orange) and no-MSA (green), both rising sharply between 60 and 85 residues to plateau near $1.0$. \textbf{(f)} Within-target pass rates across the 85-residue threshold for the four targets that straddle the boundary (VirB8, TrkA, IL10Ra, and Pdl1) under +MSA (solid) and no-MSA (dashed): VirB8 jumps from 0\% below 85 residues to 100\% above in both arms, whereas TrkA, Pdl1, and IL10Ra already fold accurately at smaller crop sizes.}
\end{figure}

Because minimal halo crops save the most compute but risk destabilizing the target pocket, we swept halo thickness across all designs to find the smallest radius that reliably preserves interface geometry. Across all targets, interface RMSD improves as the halo widens (\textbf{Fig. 3a} shows per-target $\mathrm{mean} \pm  1 \,\text{SD}$ on confirmed binders). VirB8 shows the sharpest sensitivity to halo radius: because its MSA is shallow, its median interface RMSD remains above the $3 \,\text{\AA}$ threshold until the halo reaches $\sim  9 \,\text{\AA}$, dropping from $5.73$ to $4.38$, $3.90$, $0.29$, and $0.24 \,\text{\AA}$ across halos of $0, 3, 6, 9,$ and $11 \,\text{\AA}$ (with a valid-fold fraction of $0.44$ at halo 6). IL2Ra fails only at the bare pocket ($6.54 \,\text{\AA}$ at halo 0 versus $0.27 \,\text{\AA}$ at halo 3). For IL10Ra, coevolutionary information in the MSA stabilizes the bare pocket (median $1.45 \,\text{\AA}$ with MSA versus $7.03 \,\text{\AA}$ single-sequence). Most other targets plateau below $\sim  0.6 \,\text{\AA}$ RMSD by halo 6. In practice, a $\sim  6 \,\text{\AA}$ halo provides sufficient structural context for most targets, while shallow-MSA targets benefit from a wider $\sim  9 \,\text{\AA}$ shell. Refolding the full halo sweep in single-sequence mode (no MSA) on $\mathrm{ipSAE}_{\min}$ across all five radii shows a similar trend in structural and ranking prediction: across the 11 well-powered targets, interface RMSD improves as the halo widens (\textbf{Supporting Fig. 4a}) and mean ranking rises from the bare pocket and levels off by an $\sim  8 \,\text{\AA}$ crop radius, with no-MSA matching or exceeding excised+MSA at every radius (mean AP of $0.37, 0.47, 0.48, 0.51,$ and $0.50$ at radii of $5, 8, 11, 14,$ and $16 \,\text{\AA}$, compared with $0.34, 0.43, 0.42, 0.42,$ and $0.45$ with MSA) and reaching $\sim  95\%$ of the full-target+MSA baseline (AP $0.51$ and ROC-AUC $0.79$ at $14 \,\text{\AA}$ versus $0.54$ and $0.81$; \textbf{Fig. 3b,c}, \textbf{Supporting Fig. 4b,c}). As before, the shallow-MSA target SARS-CoV-2 RBD is the lone exception that performs worse without its MSA.

A fixed radial halo captures different numbers of residues depending on local target topology. For example, across EGFR designs the median crop is already 374 residues at the smallest halo, 4 to 9 times larger than the other targets. When plotted the halo sweep against the absolute number of retained target residues for confirmed binders (pooled across the nine sweep targets) crops shorter than $\sim  85$ residues pass the $3 \,\text{\AA}$ gate 32\% of the time ($n = 41$), compared with 97\% for crops of 85--200 residues ($n = 553$) (\textbf{Fig. 3d,e}, \textbf{Supporting Fig. 4d}). Much of this pooled drop comes from VirB8, where 0\% of the 24 sub-85-residue folds pass (compared with 76\% of the 17 sub-85-residue folds across other targets, and 86\% for VirB8 above 85 residues); by contrast, confirmed binders for Pdl1 and TrkA refold accurately even on small crops (100\% below 85 residues for Pdl1, $n = 6$; 88\% for TrkA, $n = 8$; \textbf{Fig. 3e,f}). A safe crop window is therefore roughly $\sim  85$ residues; the pass rate climbs across 60--85 and plateaus near $1.0$ above 85, which generally corresponds to a $\sim  6 \,\text{\AA}$ halo. Overall, overly tight crops degrade structural accuracy before they degrade relative ranking, while keeping crop lengths moderate helps complexes stay within lower AF3 token buckets.

\begin{figure}[!htbp]
\centering
\includegraphics[width=\textwidth]{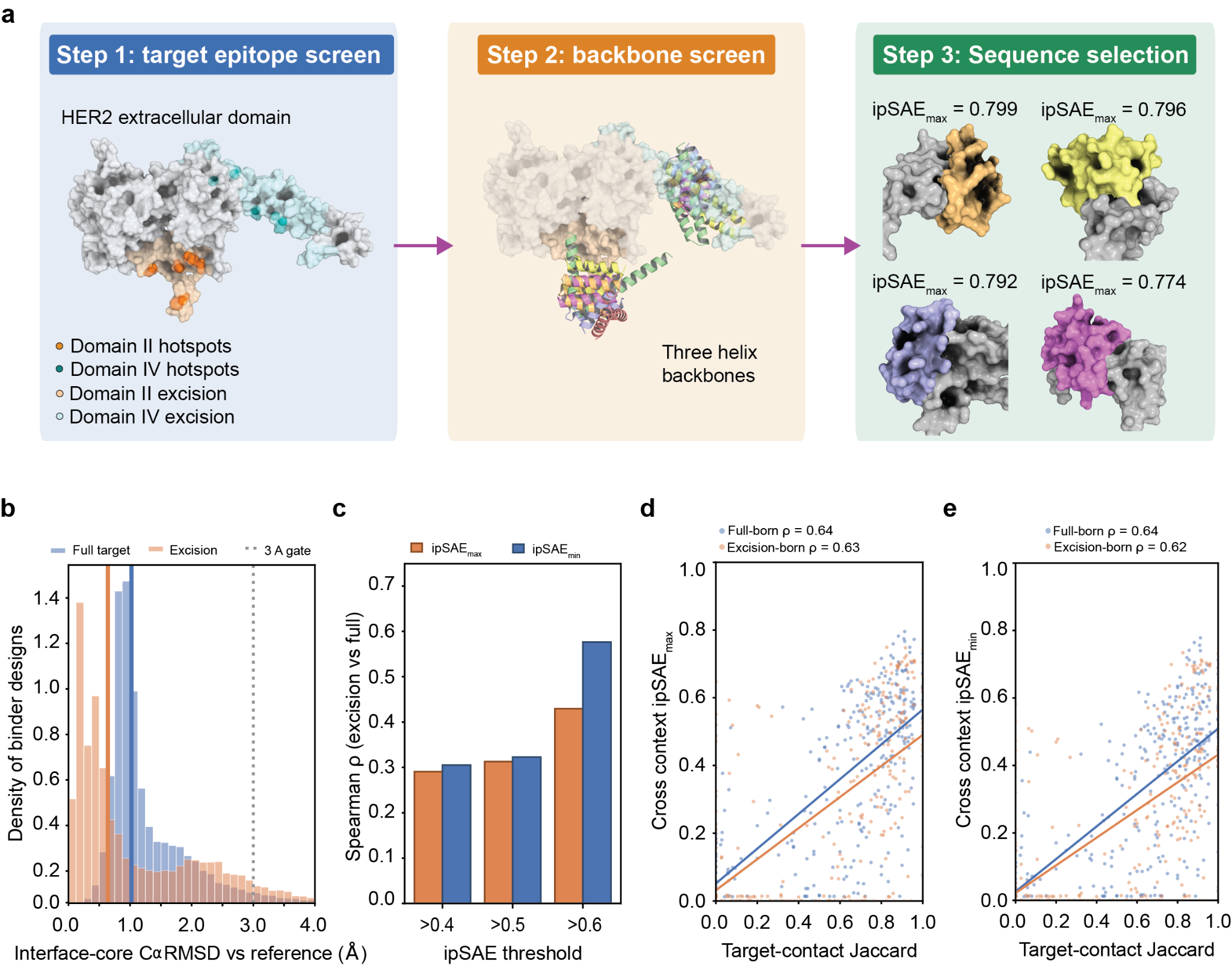}
\caption*{\textbf{Figure 4} | \textbf{The fixed epitope-excision pipeline and its cross-context fidelity (HER2).} \textbf{(a)} Schematic of the fixed end-to-end excision pipeline: target preparation (select a target epitope, excise a contiguous 85--130-residue window around it, and construct a crop MSA and inline self-template), cluster screening (enumerate candidate hotspot clusters per domain, run a mini-screen of $\sim  30$ backbones per cluster, rank by hit rate, and advance the top clusters to production), backbone generation (generate RFdiffusion three-helix-bundle "HHH" scaffolds targeted to the chosen hotspots with steric-clash filtering), \textbf{sequence design and folding} (design sequences with ProteinMPNN including a top-off pass on productive backbones, then co-fold binder and crop in AF3 using the self-template), and gating and ranking on the fixed hit criteria ($\mathrm{ipSAE}_{\max}$ $> 0.45$ and interface-core $\mathrm{C}\alpha$ RMSD $< 3.0 \,\text{\AA}$). \textbf{(b)} Interface-core $\mathrm{C}\alpha$ RMSD between AF3 refolds and design references across the HER2 production campaigns for full-domain (blue) versus excised-crop (orange) designs; both arms recover the designed interfaces accurately, with the excised arm showing a slightly tighter distribution (median $0.64 \,\text{\AA}$ versus $1.02 \,\text{\AA}$; 96\% and 91\% of designs under the $3 \,\text{\AA}$ threshold, respectively). \textbf{(c)} Filtering excised designs by higher native-context $\mathrm{ipSAE}_{\max}$ increases the rank correlation between excised and full-domain $\mathrm{ipSAE}_{\max}$ scores (Spearman $\rho $ rising from $0.29$ across all designs to $0.46$ at $\mathrm{ipSAE}_{\max}$ $< 0.7$). \textbf{(d)} Per-binder epitope conservation across contexts, measured as the Jaccard overlap of target-contact residues between the full domain and the crop, correlates with cross-context $\mathrm{ipSAE}_{\max}$ (Spearman $\rho  = +0.64$ for full-domain designs and $+0.63$ for excised-crop designs), indicating that binders preserve their confidence scores when they maintain their intended contact network.}
\end{figure}

Before applying epitope excision to a new target, we froze the end-to-end design pipeline and evaluated whether designs generated on cropped targets behave consistently when evaluated on full domains. In the fixed pipeline (\textbf{Fig. 4a}), target preparation selects an epitope, and excises a contiguous 85--130-residue window. Candidate hotspot clusters are then screened in small batches, followed by RFdiffusion three-helix-bundle ("HHH") backbone generation, ProteinMPNN sequence design, and AF3 refolding, with hits defined by a fixed threshold ($\mathrm{ipSAE}_{\max}$ and interface-core $\mathrm{C}\alpha$ RMSD $< 3.0 \,\text{\AA}$). We tested this setup on HER2, comparing campaigns run against full target domains versus excised domain crops. Across the HER2 production runs, the excised campaign recovered designed interfaces at least as accurately as full-domain folding (median interface-core $\mathrm{C}\alpha$ RMSD of $0.64 \,\text{\AA}$ versus $1.02 \,\text{\AA}$, with $> 90\%$ of designs under the $3 \,\text{\AA}$ gate in both cases; \textbf{Fig. 4b}). Interface confidence scores also tracked across contexts: filtering for higher native-context $\mathrm{ipSAE}_{\max}$ and $\mathrm{ipSAE}_{\min}$ strengthened the correlation between excised and full-domain (Spearman $\rho $ increasing from $\sim  0.3$ across all designs to $0.46$ and $0.59$ for designs with $\mathrm{ipSAE}_{\max}$, $\mathrm{ipSAE}_{\min}$ $> 0.6$; \textbf{Fig. 4c}). Furthermore, the Jaccard overlap of target-contact residues between the crop and the full domain predicted cross-context $\mathrm{ipSAE}_{\max}$ (Spearman $\rho  = +0.64, +0.63$ for full-domain and excised designs; \textbf{Fig. 4d}) and $\mathrm{ipSAE}_{\min}$ (Spearman $\rho  = +0.64, +0.63$ for full-domain and excised designs; \textbf{Fig. 4e}), confirming that candidates retain their interface confidence when their designed contact network is preserved.

\begin{figure}[!htbp]
\centering
\includegraphics[width=\textwidth,height=0.55\textheight,keepaspectratio]{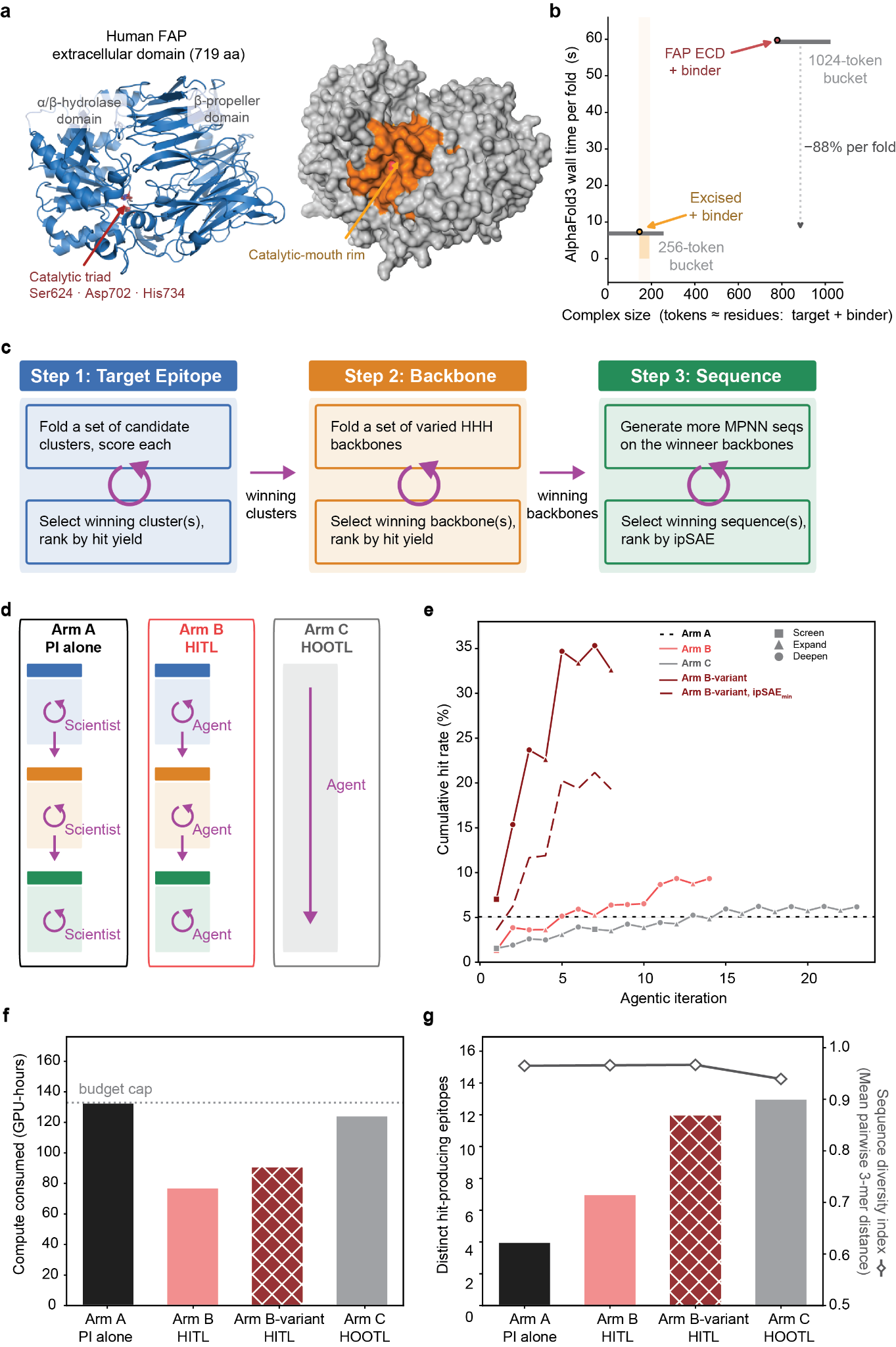}
\caption*{\textbf{Figure 5} | \textbf{The FAP autonomy experiment: target, compute economy, and autonomy ladder.} \textbf{(a)} Structure of the human FAP extracellular domain (PDB 1Z68, chain B, 719 residues), comprising an $\alpha /\beta $-hydrolase domain and a $\beta $-propeller domain (left) whose catalytic triad (Ser624, Asp702, His734; red) sits inside a recessed substrate-channel opening (right; the targeted mouth-rim epitope is highlighted in orange around the central recessed triad). \textbf{(b)} Expected AF3 compute reduction under the token-bucket model (\textbf{Fig. 1}): the intact FAP ECD plus binder ($\sim  779$ tokens) falls into the 1024-token bucket ($59.3 \,\text{s}$ per fold), whereas an 85--130-residue crop plus binder fits into the 256-token bucket ($6.9 \,\text{s}$ per fold), giving an $\sim  88\%$ per-fold reduction in runtime. \textbf{(c)} Schematic of the excision pipeline with decision points indicated with purple arrows. \textbf{(d)} Autonomy conditions: Arm A (PI alone), Arm B (human--agent hybrid, where the investigator specifies the catalytic-rim strategy and the agent controls tactical execution), and Arm C (fully autonomous agent selecting strategy, target surface, and hotspot sets) use the exact same hit definition ($\mathrm{ipSAE}_{\max}$ $> 0.45$ and interface-RMSD $< 3.0 \,\text{\AA}$) and GPU-hour budget cap. The B-variant uses the same hybrid autonomy setup as Arm B but modifies two scoring settings (disabling the target MSA and scoring on $\mathrm{ipSAE}_{\min}$). \textbf{(e)} Cumulative hit rate across agentic iterations for Arms B and C against the Arm A baseline (dotted line), with marker shapes denoting controller actions; the B-variant is plotted on both $\mathrm{ipSAE}_{\max}$ (solid) and $\mathrm{ipSAE}_{\min}$ (dashed). \textbf{(f)} Total GPU-hours consumed per condition (dotted line marks the budget cap). \textbf{(g)} Design diversity across arms: bars show the number of distinct epitopes yielding $\ge  1$ hit (rising from 4 in Arm A to 7 in Arm B, 12 in the B-variant, and 13 in Arm C), while diamonds show mean pairwise 3-mer sequence diversity ($\sim  0.94$--$0.97$).}
\end{figure}

With the excision pipeline validated on HER2, we turned to the extracellular domain of fibroblast activation protein $\alpha $ (FAP ECD; PDB 1Z68, chain B, residues 39--757; 719 residues), a serine protease whose catalytic triad (Ser624/Asp702/His734) sits at the base of a recessed substrate channel (\textbf{Fig. 5a}). Because the intact 719-residue ectodomain is too large for repeated full-structure diffusion and costly to refold at scale, we used excision to isolate contiguous 85--130-residue crops around candidate epitopes and ran the same RFdiffusion ("HHH" scaffolds), ProteinMPNN, and AF3 pipeline (\textbf{Fig. 5b}). Measured on an A100 for one epitope cluster with an 85-residue crop, excision cut RFdiffusion, Protein MPNN, and AF3 times per unit by 89\%, 66\%, and 88\% respectively (\textbf{Supporting Fig. 5a}). We note that capping crops at $\le  130$ residues is slightly more conservative than our benchmark minimum, where 130--200-residue crops achieved a 98.8\% geometry pass rate; relaxing this cap would trade a modest increase in compute for additional structural context on difficult epitopes.

Using FAP as a test case, we examined varying degrees of decision-making autonomy by an LLM agent when steering the excision pipeline (\textbf{Fig. 5c,d}). We kept the hit definition ($\mathrm{ipSAE}_{\max}$ $> 0.45$ interface-RMSD $< 3.0 \,\text{\AA}$) and the total compute budget (capped at the 132.87 GPU-hours spent by Arm A, PI alone) identical across runs, varying only how campaign decisions were made across three primary conditions: Arm A (fixed excision pipeline as in \textbf{Fig. 4a}; PI alone), Arm B (a human--agent hybrid; Human-in-the-Loop, HITL, in which an LLM agent, Gemini 3.1 Pro, uses the excision pipeline in \textbf{Fig. 5c} but chooses the specific epitope sets, crop sizes, ProteinMPNN parameters, compute allocation, and iteration decisions), and Arm C (a fully autonomous agent that selects its own targeting strategy and surface patches from scratch without human pre-filtering; Human-out-of-the-Loop, HOOTL). At each round in Arms B and C, Gemini 3.1 Pro receives a structured summary of per-cluster hit counts, remaining GPU budget, and current generation settings and selects a single typed action (propose new epitope or hotspot clusters, iteration/backtrack, epitope/backbone/sequence expansion). We also ran a fourth condition, the B-variant, which uses the same hybrid autonomy setup as Arm B but modifies two scoring settings (turning the target MSA off and ranking on $\mathrm{ipSAE}_{\min}$) to separate the effect of the scoring setup from the effect of agent autonomy.

\begin{table}[!htbp]
\centering
\caption*{\textbf{Table 1} $\vert$ \textbf{Autonomy-ladder outcomes.}}
\resizebox{\textwidth}{!}{%
\begin{tabular}{llrrrrrr}
\toprule
\textbf{Arm} & \textbf{Decision-maker} & \textbf{Hits / designs} & \textbf{Hit rate} & \textbf{GPU-hours} & \textbf{Efficiency} & \textbf{Epitopes} & \textbf{Seq.\ div.} \\
\midrule
A (PI alone) & human & 1,672 / 33,024 & 5.06\% & 132.87 & 12.6 & 4 & 0.964 \\
B (hybrid) & human + agent & 1,790 / 19,184 & \textbf{9.33\%} & 77.18 & \textbf{23.2} & 7 & 0.965 \\
C (agent alone)* & agent & 1,803 / 29,176 & 6.18\%* & 124.43 ($\sim 94\%$) & 14.5 & \textbf{13} & 0.939 \\
B-variant\textsuperscript{$\dagger$} (apparatus ablated) & human + agent & 7,106 / 21,784 & 32.62\%\textsuperscript{$\dagger$} & 90.94 & 78.1\textsuperscript{$\dagger$} & 12 & 0.966 \\
\bottomrule
\end{tabular}%
}
\vspace{4pt}
\noindent{\footnotesize * Arm C ran to completion: the agent terminated the campaign on its own decision at $124.43$ of the $132.87$ GPU-h cap ($\sim 94\%$), leaving the residual budget unspent, so its values are final.\\
\textsuperscript{$\dagger$} The B-variant sits at the same autonomy level as Arm B; it is a control that ablates two otherwise-frozen apparatus elements (target-MSA off; native ruler $= \mathrm{ipSAE}_{\min}$). Its headline rate (32.62\%, efficiency 78.1) is on the frozen $\mathrm{ipSAE}_{\max}$ ruler for comparability; on its native $\mathrm{ipSAE}_{\min}$ ruler the same 21,784 designs give 4,206 hits $= 19.31\%$ (efficiency 46.3). The B-variant is therefore excluded from the autonomy comparison and reported only to separate apparatus effects from autonomy effects.}
\end{table}

When evaluated under the shared scoring criteria, both HITL (Arm B) and HOOTL (Arm C) outperformed the PI-alone baseline (\textbf{Table 1}; \textbf{Fig. 5e}), with Arm B generating the highest hit rate (9.33\%, +84\% over the 5.06\% PI-alone baseline in Arm A) while using $\sim  58\%$ of the GPU budget (\textbf{Fig. 5f}), corresponding to $23.2$ versus $12.6$ hits per GPU-hour. The ordering was unchanged when every arm was cut off at Arm B's 77.2 GPU-h (\textbf{Supporting Fig. 5b}). Thus, having the human investigator define the biological target region while delegating tactical parameter tuning and resource allocation to the agent worked better than either manual execution or unconstrained agent search. In both agent-driven arms, cumulative hit rates climbed over successive rounds as the controller shifted compute toward productive hotspot clusters through "expand" and "deepen" actions (in Arm B, 1 initial screen, 3 expansions, and 10 deep-dives across 14 iterations). Increasing agent autonomy also widened the number of distinct hit-producing epitopes from 4 (Arm A) to 7 (Arm B) and 13 (Arm C), whereas sequence diversity among hits remained comparable across all conditions (\textbf{Fig. 5g}, \textbf{Table 1}). Without a human prior to narrow the initial search space, the fully autonomous agent spread its GPU budget across more surface patches, yielding broader epitope coverage but a lower overall hit rate than the hybrid arm. Finally, the B-variant illustrates how sensitive raw hit rates are to refolding settings: evaluated on the standard $\mathrm{ipSAE}_{\max}$ threshold, it appears to reach a 32.62\% hit rate (and 19.31\% on $\mathrm{ipSAE}_{\min}$ metric).

Inspecting the logs from the fully autonomous run (Arm C) shows that the agent arrived at a sensible targeting strategy on its own. Provided only with a strategy-neutral list of whole-target surface residues, the Arm C agent opened by proposing a mouth-occlusion strategy aimed at well-ordered, solvent-exposed residues along the rim of the catalytic channel within $15 \,\text{\AA}$ of the catalytic triad ($d_\mathrm{cat} < 15 \,\text{\AA}$), closely matching the steric-blockade rationale designed by the human investigator for Arm A. The agent then filtered candidate hotspots by solvent-accessible surface area ($\mathrm{SASA} > 60$), crystallographic order (B-factor ratio $< 1.0$), and triad distance, effectively piecing together the same buildability criteria used by human designers without being prompted with those rules. Because the agent's default structural priors already align with human target selection, constraining its search space in the hybrid arm (Arm B-variant) lets it spend its entire compute budget on productive local optimization rather than initial surface exploration.

\begin{figure}[!htbp]
\centering
\includegraphics[width=0.73\textwidth]{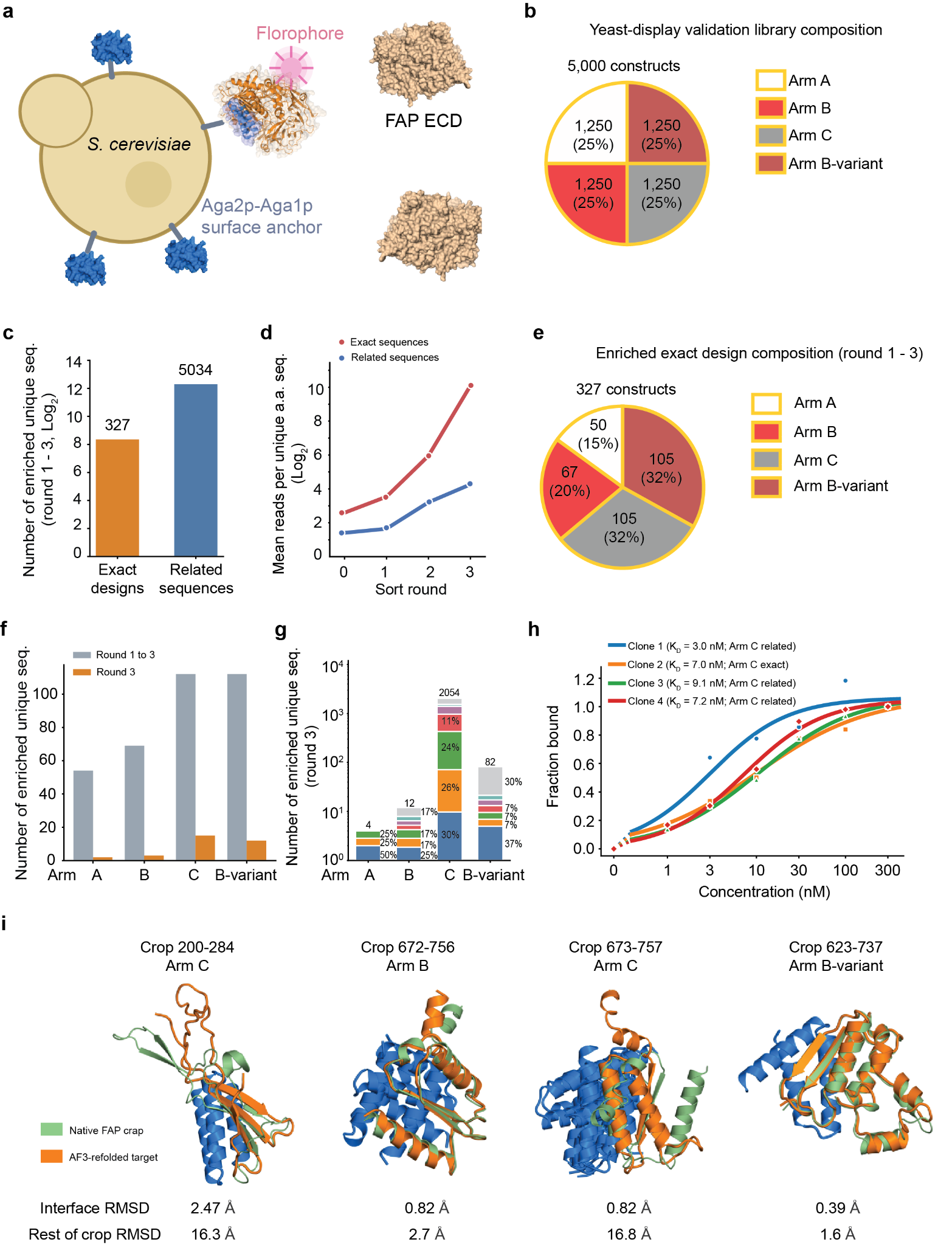}
\caption*{\textbf{Figure 6} | \textbf{Wet-lab validation of \textit{de novo} FAP binders by yeast-surface display.} \textbf{(a)} Yeast-display screening format: \textit{de novo} HHH mini-binders are expressed on the \textit{S. cerevisiae} surface as Aga2p--Aga1p fusion proteins and interrogated with soluble, biotin-conjugated FAP ectodomain. Cells displaying functional binders bind target protein and are isolated by fluorescence-activated cell sorting (FACS). \textbf{(b)} Composition of the 5,000-member yeast-display oligonucleotide synthesis library, containing 1,250 designs per campaign arm (Arm A, Arm B, Arm C, and Arm B-variant). \textbf{(c)} Number of enriched unique binder sequences detected, pooled across arms and across FACS rounds 1--3. Exact designs are sequences that are amino-acid-identical to a design and related sequences are those that differ from the designed sequences by 1--14 amino acids. \textbf{(d)} Mean read count per unique amino acid sequence for exact and related sequences across sort rounds 0--3. \textbf{(e)} Distribution of the 327 enriched exact designs across campaign arms. \textbf{(f)} Number of unique exact designs recovered per arm in rounds 1--3 or round 3 only. \textbf{(g)} Number of unique related sequences recovered in round 3 per arm, on a log scale, with segments showing the share from each parent design. \textbf{(h)} Equilibrium binding titrations for four representative enriched clones, displaying single-digit nanomolar apparent equilibrium dissociation constants ($K_D = 3.0$--$9.1 \,\text{nM}$). \textbf{(i)} AF3-refolded excised crop compared with the native FAP crop, superposed at the binding interface.}
\end{figure}

To validate these computational predictions experimentally, top-ranked designs across conditions were evaluated using yeast-surface display against the FAP ectodomain (\textbf{Fig. 6a}). Candidates were expressed on \textit{S. cerevisiae} via an Aga2p--Aga1p anchor. We constructed a 5,000-member oligonucleotide library containing 1,250 designs per campaign arm (Arms A, B, C, and B-variant; \textbf{Fig. 6b}), augmented with random mutagenesis (2--3 nucleotide substitutions per sequence) to explore local sequence space. The library was screened with fluorophore-conjugated target protein across three sequential rounds of fluorescence-activated cell sorting (FACS) at decreasing target concentrations ($300 \,\text{nM}$, $150 \,\text{nM}$, and $75 \,\text{nM}$). Illumina sequencing identified 5,361 enriched unique sequences, including 327 exact matches to computationally generated designs and 5,034 related sequences within 14 amino acids of a design (\textbf{Fig. 6c}). Exact designs gained read depth far faster than related sequences across rounds (mean reads per unique sequence rising from 3 to 547 for exact designs versus $1.3$ to $9.6$ for related sequences between rounds 0 and 3; \textbf{Fig. 6d}). The enriched exact designs comprised 15\% from Arm A ($n = 50$), 20\% from Arm B ($n = 67$), 32\% from Arm C ($n = 105$), and 32\% from Arm B-variant ($n = 105$) as shown in \textbf{Fig. 6e}. Arms C and B-variant recovered more exact designs than Arms A and B, both across rounds 1--3 and in round 3 alone (\textbf{Fig. 6f}). Related sequences enriched in round 3 were concentrated in a few parent designs (\textbf{Fig. 6g}). Clonal binding assays on four representative enriched leads demonstrated nanomolar binding affinities ranging from $3.0 \,\text{nM}$ to $9.1 \,\text{nM}$ $K_D$ (\textbf{Fig. 6h}). AF3 refolds of the top exact designs reproduced the native FAP interface while the regions outside the interface deviated substantially more (\textbf{Fig. 6i}).

\section*{Discussion}

This study examines two practical ways to improve the efficiency of \textit{de novo} binder campaigns against large protein targets. First, cropping targets to local epitope windows substantially reduces AlphaFold3 refolding costs while preserving both interface geometry (median RMSD of $0.37 \,\text{\AA}$) and binder ranking. With the target MSA retained, excised ranking dips on a few shallow-MSA or fragile-pocket targets (Pdl1, IL7Ra); single-sequence refolding recovers these, leaving SARS-CoV-2 RBD, which depends on its coevolutionary signal, as the lone target that performs worse when cropped (1 of 11) \cite{ref10}. Because AF3 runtime scales in discrete token buckets of 256, the compute savings from excision depend on target size: small proteins that already fit in the 256-token bucket see no speedup, whereas large multi-domain targets drop into much cheaper buckets. The efficiency gains during refolding should apply similar scaling advantages to upstream backbone generation steps, such as RFdiffusion, that scale quadratically with sequence length. Our halo-sweep experiments show that the contextual information about the binding interface does affect structural and ranking accuracy while preserving structural floor of $\sim  85$ retained target residues, making it a promising way to tackle large receptors, viral glycoproteins, and enzymes that are otherwise too expensive to fold repeatedly at full length.

Second, pairing an excision-based pipeline with an LLM agent improves computational yield under a fixed GPU budget. Across our controlled comparison on targeting FAP, the human--agent hybrid (Arm B) achieved the highest hit rate at the lowest compute cost. This result points to a practical division of labor: having a human researcher specify the general framework for the campaign strategy, such as ranking metrics and structural properties or biological target regions, while letting the LLM agent handle iterative parameter selection, crop sizing, and budget allocation across different stages improves throughput over manual management. Notably, when left to choose its own strategy from a whole-target residue map, the fully autonomous agent (Arm C) independently converged on the same catalytic-mouth occlusion strategy and surface-filtering criteria used by the human investigator. Even so, under a capped compute budget, giving the agent a human strategic prior proved more efficient than having it spend GPU hours discovering where to bind, whereas full autonomy was useful for spreading designs across a wider set of 13 distinct epitopes.

To evaluate these computational predictions at scale, wet-lab validation of the \textit{de novo} FAP binders was conducted using yeast-surface display. Among the enriched exact sequences, the higher representation of Arms B-variant and C relative to Arms A and B is consistent with the \textit{in silico} metrics. The improved representation in the Arm B-variant indicates that enhanced ipSAE in this arm may rescue expression. Furthermore, higher representation of the Arm B-variant relative to Arm B suggests that excluding the target MSA does not impair binder prediction, consistent with our \textit{in silico} hit rates observed in another recent study \cite{ref15}. Finally, the top four clones isolated after three rounds of sorting exhibited single-digit nanomolar apparent $K_D$ values, indicating physiologically relevant and specific interactions with the disease target FAP. Future work will include the binding characterization of solubly expressed binders, evaluation of FAP domain specificity, and inhibition assays of FAP catalytic activity.

Several caveats apply to these findings. Refolding excised crops without a target MSA recovers ranking accuracy on targets with deep alignments but hurts performance on a shallow-MSA target SARS-CoV-2 RBD, suggesting that single-sequence refolding may not apply to all targets. Second, in the FAP autonomy experiment, the fully autonomous agent in Arm C chose to stop at $\sim  94\%$ of the shared compute budget; while its final numbers reflect the agent's own stopping decision, a policy tuned to exhaust the remaining 6\% of the budget might have yielded additional hits. Finally, our autonomy comparison focused on a single large enzyme target (FAP) and a single backbone class (three-helix bundles) driven by one LLM controller; testing how these autonomy trade-offs generalize to other target classes remains future work.

Overall, our results outline a straightforward workflow for large macromolecular targets: crop the target to a local 85--130-residue window (or $\sim  6 \,\text{\AA}$ halo), score candidates with a fixed interface-confidence metric, and use an LLM agent to manage the iterative design loop within a human-selected target region. Applying this combination turned the 719-residue FAP ectodomain into a scalable computational design campaign while substantially lowering GPU usage.

\section*{Methods}

\subsection*{Benchmark dataset}

All screening analyses use the 15-target \textit{de novo} binder benchmark of Overath et al \cite{ref12}, published as 3,766 experimentally characterized binders. From the public release we retained designs with complete sequence records and restored recoverable peptide-MHC (pMHC) designs, giving 3,760 designs used for the ranking analysis (\textbf{Fig. 1}, \textbf{Supporting Table 1}). The excision-fidelity analysis (\textbf{Fig. 2}) is restricted to the 3,666 designs for which a single contiguous interface window is defined; it excludes 94 pMHC designs whose epitope sits within the MHC groove and therefore has no compact crop.

\subsection*{Target excision}

For each design, target residues with any atom within the contact radius of the binder were retained together with all target residues within an additional halo shell ($\mathrm{keep-radius} = 5 \,\text{\AA} \,\text{contact} + \mathrm{halo}$; halos $0/3/6/9/11 = \mathrm{radii} \,5/8/11/14/16 \,\text{\AA}$); the pocket was folded in place of the intact target. The crop is the pure contiguous span of the primary interface chain (no growing or capping in the benchmark reconstruction). Halo $11 \,\text{\AA}$ is the shipped operating point; halo $\approx  6 \,\text{\AA}$ is the safe floor identified here ($\approx  9 \,\text{\AA}$ for shallow-MSA targets).

\subsection*{Folding, scoring, and metrics}

Complexes were folded with AlphaFold3 (on-demand memory, \texttt{--norun\_data\_pipeline}, default token buckets). Ranking used $\mathrm{ipSAE}_{\max}$. Full-target references used a deep MSA and full-target template; all excised folds in this reconstruction used a per-excision MSA sliced offline from the cached full-target \texttt{a3m}. Interface geometry $= $ target interface-core $\mathrm{C}\alpha$ RMSD between re-fold and design reference (validity gate $3 \,\text{\AA}$); for geometry-vs-halo curves the median is taken over all designs so failed folds remain visible, with validity fractions reported separately. Ranking accuracy $= $ AP and ROC-AUC of $\mathrm{ipSAE}_{\max}$, with per-target values as the replication unit ($n_\mathrm{pos} \ge  5$ for informative targets); ranking error bars use 2,000-resample design bootstraps, and full-vs-excised deltas use a per-target paired bootstrap. Baselines in the halo-sweep analyses (\textbf{Fig. 3}, \textbf{Supporting Fig. 4}) are computed on the halo-matched design subset, the designs folded at every crop radius, so curves are comparable across radii; the all-designs baselines used elsewhere (\textbf{Fig. 2}, \textbf{Supporting Fig. 3}) are therefore marginally lower (full-target+MSA AP $0.47$ vs $0.49$ for $\mathrm{ipSAE}_{\max}$, $0.51$ vs $0.54$ for $\mathrm{ipSAE}_{\min}$). AF3 runtime was modeled as $\mathrm{bucket} = ceil(\mathrm{tokens}/256) \times  256$ (minimum 256), with per-fold times $6.9 / 17.5 / 35.0 / 59.3 \,\text{s}$ for the $256 / 512 / 768 / 1024$ buckets; batch saving per target $= (sum \mathrm{full-target time} - sum \mathrm{excised time}) / sum \mathrm{full-target time}$.

\subsection*{FAP target and excision}

The target is the FAP extracellular domain (PDB 1Z68, chain B, residues 39--757; $\sim  719$ residues), catalytic triad Ser624/Asp702/His734 lining a near-buried substrate-channel mouth. The pipeline excises a compact continuous crop of 85--130 residues around the chosen epitope; excision is frozen as a mechanism across all arms, with crop size the only tunable knob (a direct compute lever).

\subsection*{Design toolchain}

Each candidate epitope is turned into binders by a fixed pipeline: RFdiffusion (three-helix-bundle "HHH" scaffold family only), ProteinMPNN sequence design, AlphaFold3 refolding, followed by refold-based interface evaluation. Binder length is a free knob hard-capped at $\le  71$ residues. Default generation policy: ProteinMPNN temperatures $\{0.1, 0.3\} \times  4$ batches, target MSA on, binder MSA off; the agentic arms may adjust these.

\subsection*{Frozen hit metric and budget}

A design is a hit iff $\mathrm{ipSAE}_{\max}$ $> 0.45$ AND target-interface RMSD $< 3.0 \,\text{\AA}$ (interface RMSD on a side-chain-bearing refold of the interface core). No arm may redefine a hit; robustness/$n_\mathrm{core}$ and functional-occlusion diagnostics are computed \textit{post hoc} and never enter any decision or hit count. All arms share one currency, measured GPU-hours, capped at the PI campaign's total ($132.87$ GPU-h); GPU-hours were timed per subprocess and the controller refuses any action whose estimated cost would exceed the remaining budget. Design count is not comparable across arms (per-design cost varies with crop and binder length) and is reported for reference only. Campaigns ended when the agent chose to stop or when the next action exceeded the remaining budget: Arm B halted on the agent's own action after 19,184 designs ($77.18$ GPU-h); Arm C stopped by agent decision at $124.43$ GPU-h (94\%); the B-variant ended when its next planned action exceeded the GPU-h remaining.

\subsection*{Agentic controller}

Arm A is the PI-alone campaign using the fixed pipeline. A closed-loop controller wraps the frozen pipeline. Each iteration it builds a structured observation (per-cluster hit counts, budget remaining, current generation policy, never $n_\mathrm{core}$, robustness, or buildability marks) and an LLM agent (Gemini 3.1 Pro) returns one typed action \cite{ref16}: \texttt{propose\_strategy} (required first move), \texttt{propose\_clusters}, \texttt{configure} (GPU-free), \texttt{screen} / \texttt{expand} / \texttt{deepen} / \texttt{adjust\_crop} / \texttt{backtrack} / \texttt{stop}. In Arm B the PI fixes the strategy and the agent registers only consistent epitope tactics; in Arm C the agent reads a strategy-neutral whole-target surface map, declares its own strategy, and registers hotspot sets of any arity anywhere with no buildability pre-filter (every screened set costs GPU-hours whether or not buildable). The controller is hardened against transient API failures (exponential-backoff retries, bounded timeouts). The Arm B-variant repeats Arm B with the target MSA off and hits ranked in $\mathrm{ipSAE}_{\min}$.

\subsection*{Compute benchmarking and matched-budget analysis}

To measure the compute saving directly, we ran RFdiffusion, Protein MPNN and AF3 against either the full 719-residue FAP ectodomain or an 85-residue excised crop for one epitope cluster, on a single A100 40 GB GPU. RFdiffusion time per binder was taken from the difference between 10 single-binder calls and 5 calls of 10 binders for the crop and 3 for full FAP, which removes fixed startup time. ProteinMPNN was timed per run of 3 backbones. AF3 was timed per fold on 12 identical binder sequences per condition, excluding the first fold, with no target MSA. For the matched-budget comparison, each arm's hits were counted up to $77.2$ GPU-h, the budget used by the shortest arm (Arm B).

\subsection*{Diversity metrics}

Epitope breadth $= $ number of distinct hit-producing epitopes (unique hotspot-set clusters with $\ge  1$ hit). Sequence diversity $= $ mean pairwise 3-mer Jaccard distance over hit binder sequences ($1.0 =$ maximally diverse), on a capped random sample per arm. Backbone breadth $= $ distinct $(\mathrm{epitope}, \mathrm{backbone})$ pairs producing hits.

\subsection*{Yeast display}

Yeast surface display screening and binding characterization was performed as described previously by Boder and Wittrup \cite{ref17} in-house and with DuneX Biosciences. Briefly, designed protein sequences were codon-optimized for recombinant expression in microbial hosts. For display on the \textit{S. cerevisiae} cell surface as Aga2p--Aga1p fusion proteins, coding sequences were extended with flanking regions homologous to the display vector to facilitate homologous recombination in \textit{S. cerevisiae}. Following confirmation of expression, yeast libraries displaying the constructs were enriched for binding to biotinylated FAP through three sequential rounds of fluorescence-activated cell sorting (FACS). The library comprised 5,000 oligonucleotides (1,250 designs per arm) with random nucleotide substitutions per sequence, sorted at 300, 150, and $75 \,\text{nM}$ FAP. Reads were translated and matched to the designs: "exact" sequences are amino-acid-identical to a design, "related" sequences differ from a design by 1--14 amino acids. To prevent non-specific enrichment against secondary detection reagents, the first sorting round utilized fluorophore-labeled NeutrAvidin, whereas fluorophore-labeled streptavidin was used in subsequent rounds. Sequences also detected in the streptavidin-only control were removed. For apparent $K_D$ determination, individual yeast clones isolated from the third round of screening were incubated with varying concentrations of biotinylated FAP prior to quantification of the cell-surface binding signal.

\section*{Acknowledgements}

We thank all the members of Team Morphogenesis for their help and support throughout this study. We wish to thank Ronit Levavi Morad, Bill Jia and James Manyika for leadership support. We thank Pooja Rao, Andrew Carroll and John Platt for reviews and feedback, Anthony Phalen for assistance on partnerships, Anna Sard for assistance on finance and Amanie Brik for assistance on all the things to make a lab run fast.

\section*{Competing interests}

A.M., L.C., J.R., A.H., Y.M., and A.P. are employees of Google LLC and may own Alphabet Inc. stock as part of the standard compensation package. The authors declare no other competing interests.

\section*{Author contributions}

J.W.K., L.C., and A.P. conceived the study. J.W.K., L.C., and A.P. designed the research and developed the epitope excision framework and agentic binder design workflows. J.W.K. and A.M. executed the computational benchmarking and data analysis. J.W.K. and O.D. coordinated and conducted wet-lab experiments. J.R., A.H. and Y.M. provided research leadership and strategic guidance. J.W.K. and A.P. wrote the manuscript, with input from all authors. J.W.K. and A.P. co-led the work.

\clearpage

\section*{Supporting Information}

\begin{figure}[!htbp]
\centering
\includegraphics[width=0.92\textwidth]{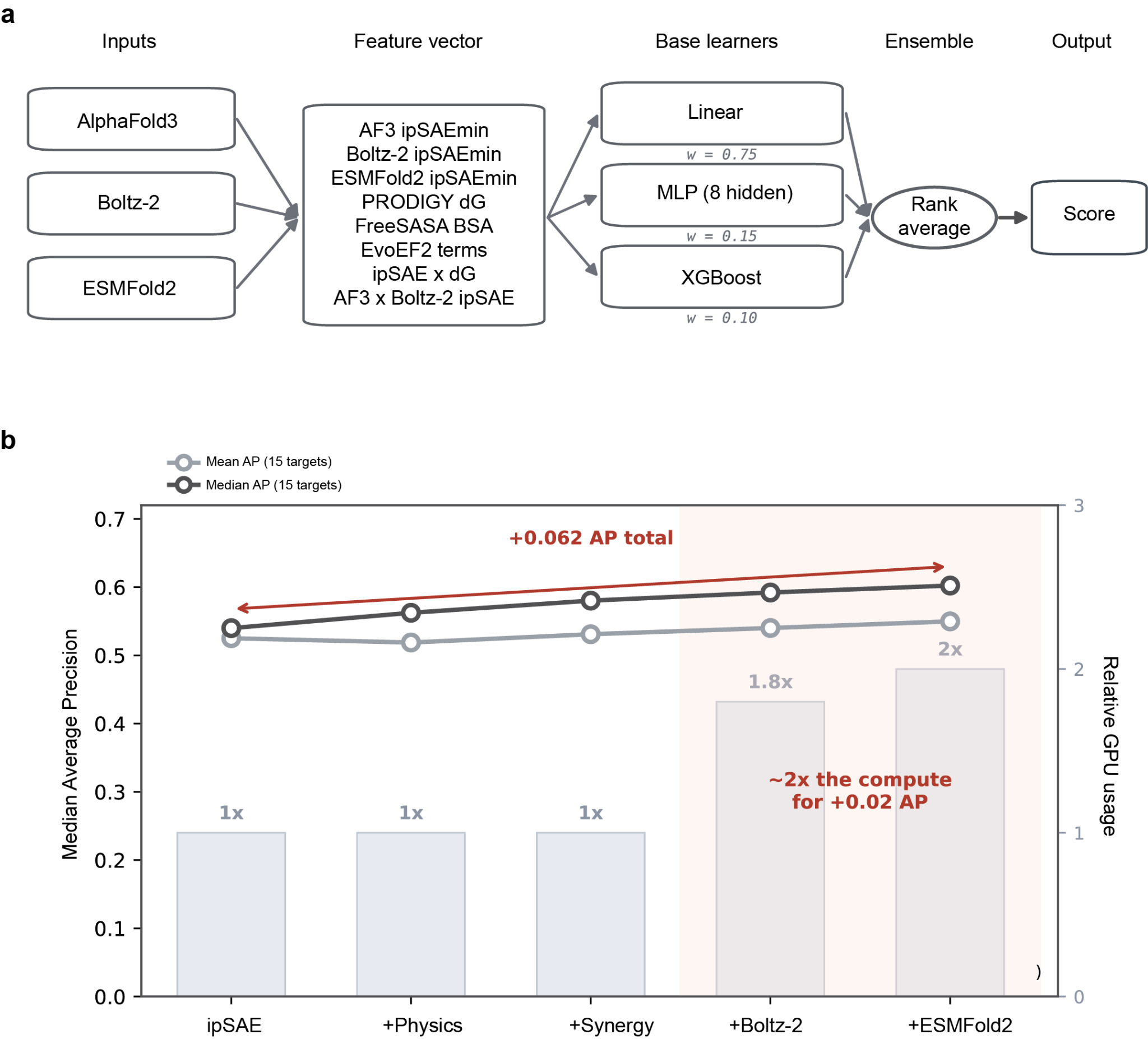}
\caption*{\textbf{Supporting Figure 1 (related to Figure 1)} | \textbf{Computational screening reaches an accuracy ceiling at rapidly rising compute.} \textbf{(a)} Architecture of the multi-engine consensus screening model. Three structure-prediction models (AlphaFold3, Boltz-2, and ESMFold2) independently fold each binder--target complex; biophysical terms (PRODIGY interface $\Delta  G$, FreeSASA buried surface area, and EvoEF2 energy terms) and interface-confidence scores are combined into a nine-element feature vector along with two interaction features ($\mathrm{ipSAE} \times  \Delta  G$ and the $\mathrm{AF3} \times  \mathrm{Boltz}$ consensus product). Features are evaluated by an L1-regularized linear model, an 8-node MLP, and gradient-boosted trees, whose rank-percentiles are combined in a fixed $0.75 / 0.15 / 0.10$ blend. Relative to AlphaFold3 ($1.0$), Boltz-2 requires $\sim  0.8 \times $ GPU compute per fold and single-sequence ESMFold2 requires $\sim  0.2 \times $. CPU biophysical calculations and ensemble inference are negligible. \textbf{(b)} Screening precision (median AP across the 15 leave-one-target-out folds; mean AP in grey) across successively more complex model architectures. Adding biophysical features and the interaction blend (which require no extra folding) yields the largest improvements ($+0.022$ and $+0.018$ AP), whereas adding Boltz-2 and ESMFold2 folds (grey bars, right axis showing relative GPU usage) adds only $+0.012$ and $+0.010$ AP while roughly doubling total GPU cost.}
\end{figure}

\begin{table}[!htbp]
\centering
\small
\setlength{\tabcolsep}{4.5pt}
\caption*{\textbf{Supporting Table 1 (related to Figure 1)} $\vert$ \textbf{Screening precision versus compute across an architecture progression, on the 15-target Overath/Cao benchmark \cite{ref12,ref13} under leave-one-group-out cross-validation (3,760 designs).}}
\begin{tabularx}{\textwidth}{l>{\raggedright\arraybackslash}X>{\raggedright\arraybackslash}p{2.7cm}rrr}
\toprule
\textbf{Configuration} & \textbf{Added component} & \textbf{Fold engines (relative GPU usage)} & \textbf{Median AP} & \textbf{Mean AP $\pm$ SD} & \textbf{$\Delta$ median AP} \\
\midrule
\textbf{AF3 $\mathrm{ipSAE}_{\min}$ alone} & raw AF3 interface score (no learning) & AF3 ($1\times$) & 0.5399 & $0.5251 \pm 0.256$ & --- \\
\textbf{Arch 1} & + physics (PRODIGY, FreeSASA, EvoEF2), L1-linear & AF3 ($1\times$) & 0.5624 & $0.5188 \pm 0.231$ & +0.0225 \\
\textbf{Arch 3A} & + synergy term ($\mathrm{ipSAE} \times \Delta G$), tri-blend 75/15/10 & AF3 ($1\times$) & 0.5802 & $0.5310 \pm 0.239$ & +0.0178 \\
\textbf{Arch 4B} & + Boltz-2 consensus ($\mathrm{AF3} \times \mathrm{Boltz}$) & AF3 + Boltz-2 ($\sim 1.8\times$) & 0.5921 & $0.5401 \pm 0.241$ & +0.0119 \\
\textbf{Arch 5} & + ESMFold2 metrics & AF3 + Boltz-2 + ESMFold2 ($\sim 2\times$) & \textbf{0.6023} & $0.5498 \pm 0.240$ & +0.0102 \\
\bottomrule
\end{tabularx}
\end{table}

\clearpage

\begin{figure}[!htbp]
\centering
\includegraphics[width=\textwidth]{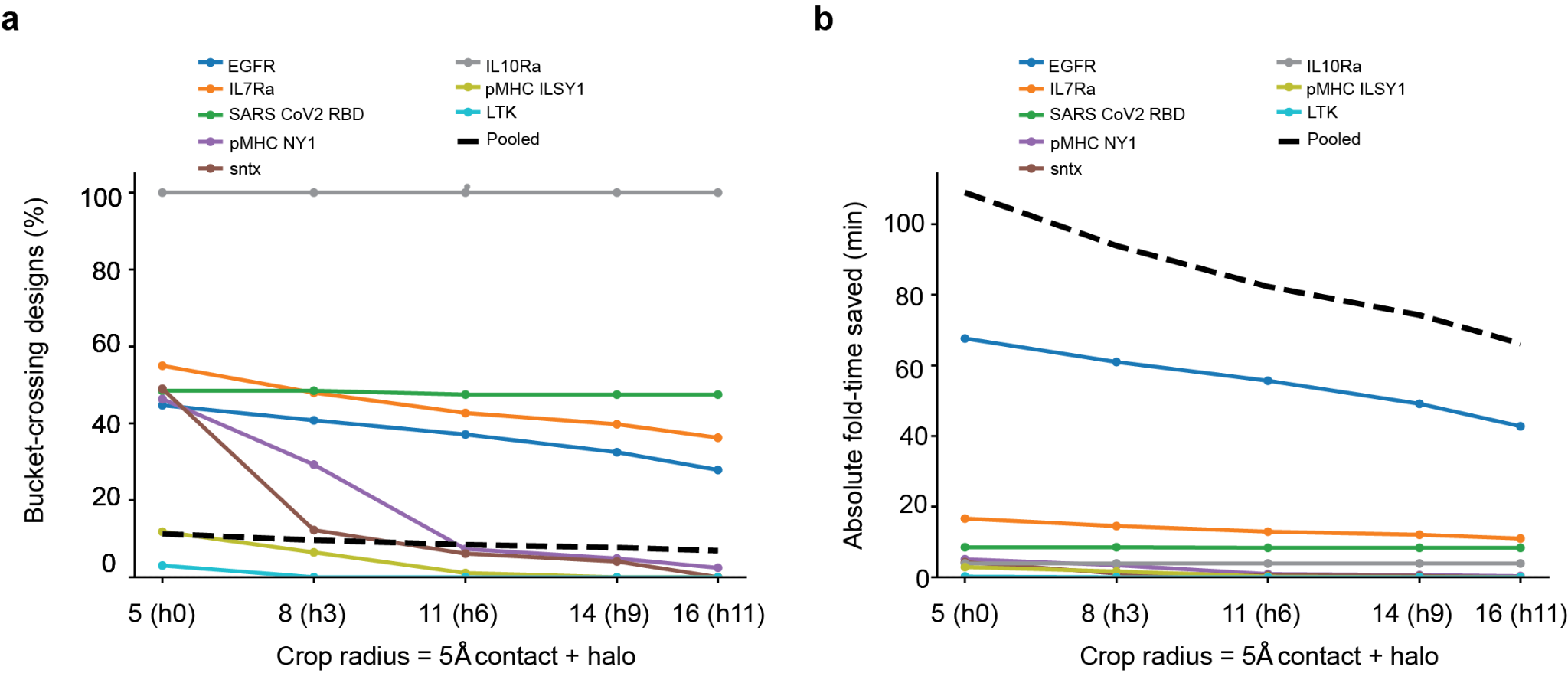}
\caption*{\textbf{Supporting Figure 2 (related to Figure 1)} | \textbf{Per-target compute economics across the crop-radius sweep.} \textbf{(a)} Percentage of each target's designs whose excised crop crosses an AF3 token-bucket boundary (bucket-crossing rate) as a function of crop radius ($5 \,\text{\AA}$ contact shell + halo, for halos $0, 3, 6, 9,$ and $11 \,\text{\AA}$), showing individual curves for saving targets and the pooled 15-target average (dashed). \textbf{(b)} Total batch AF3 fold-time saved (in minutes) versus crop radius per target. EGFR accounts for the largest wall-clock savings because its intact complex starts in a high token bucket, recovering more seconds each time a design crosses a boundary (see \textbf{Figure 1e}). Together, panels \textbf{(a)} and \textbf{(b)} separate the batch savings in \textbf{Figure 1d--e} into bucket-crossing frequency and wall-clock time recovered per crossing.}
\end{figure}

\begin{figure}[!htbp]
\centering
\includegraphics[width=\textwidth]{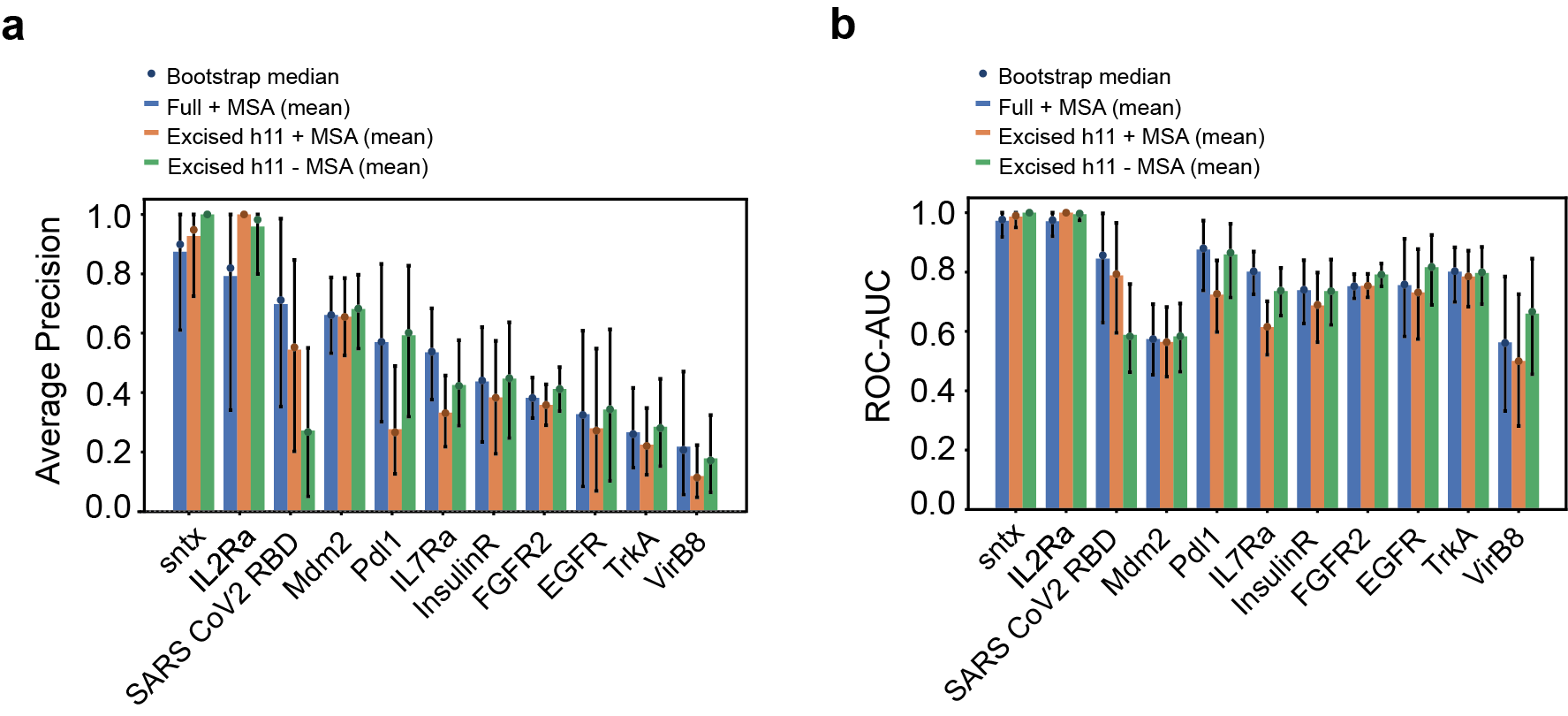}
\caption*{\textbf{Supporting Figure 3 (related to Figure 2)} | \textbf{$\mathrm{ipSAE}_{\min}$ variant of Figure 2d,e (supporting Figure 3).} \textbf{(a)} Per-target average precision (AP) and \textbf{(b)} ROC-AUC scored using $\mathrm{ipSAE}_{\min}$ (the conservative minimum across the two asymmetric chain-pair directions, as used in the source benchmark) on the informative target subset (number of binders $\ge  5$). All three arms (full-target+MSA, excised halo-11+MSA, and excised single-sequence no-MSA) show the same relative behavior as on $\mathrm{ipSAE}_{\max}$, confirming that the benchmark conclusions do not depend on chain-pair aggregation direction. (As in \textbf{Figure 2}, VirB8's full-target bar shows the reproduced full-target refold scored.)}
\end{figure}

\clearpage

\begin{figure}[!htbp]
\centering
\includegraphics[width=\textwidth]{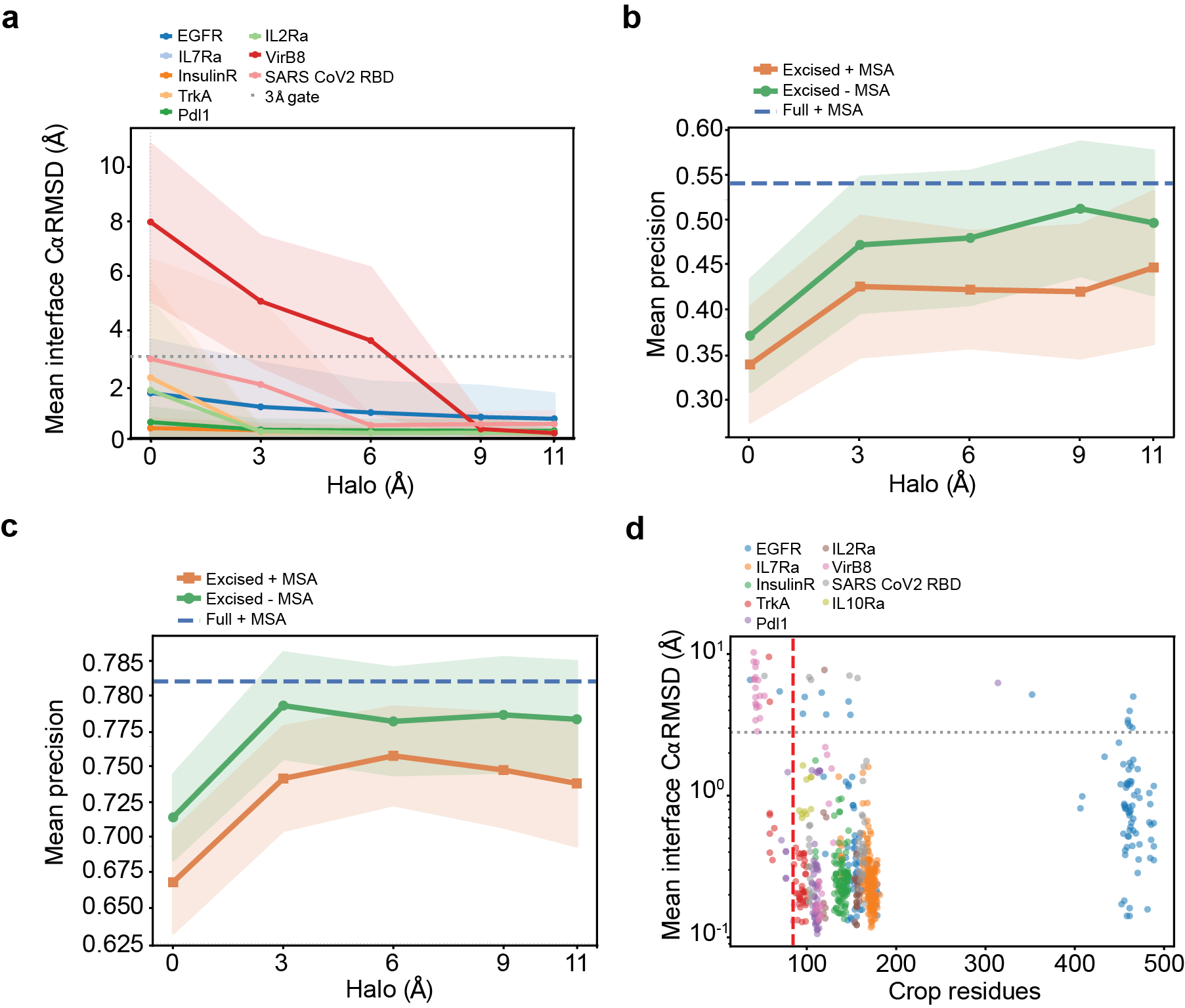}
\caption*{\textbf{Supporting Figure 4 (related to Figure 3)} | \textbf{(a)} Interface $\mathrm{C}\alpha$ RMSD versus halo radius for confirmed binders refolded in single-sequence mode (no MSA), showing per-target $\mathrm{mean} \pm  1 \,\text{SD}$ (dotted line marks the $3 \,\text{\AA}$ gate). Omitting the MSA yields the same halo-dependent improvement seen in the +MSA condition (\textbf{Fig. 3}), with VirB8 again requiring the widest halo to stabilize. \textbf{(b,c)} Ranking accuracy versus crop radius scored on $\mathrm{ipSAE}_{\min}$ across the same three arms and design-for-design matching as \textbf{Fig. 3b,c}: mean AP \textbf{(b)} and ROC-AUC \textbf{(c)} against the full-target+MSA baseline (AP $0.54$, ROC-AUC $0.81$). As with $\mathrm{ipSAE}_{\max}$, the no-MSA arm matches or exceeds excised+MSA and saturates by an $\sim  8 \,\text{\AA}$ radius. \textbf{(d)} Single-sequence counterpart to \textbf{Fig. 3d}, plotting per-fold interface $\mathrm{C}\alpha$ RMSD (log scale) against retained residue count for confirmed binders refolded without an MSA (red band marks the $> 3 \,\text{\AA}$ failure zone; dashed red line marks 85 residues), showing the same $\sim  85$-residue structural floor.}
\end{figure}

\begin{figure}[!htbp]
\centering
\includegraphics[width=\textwidth]{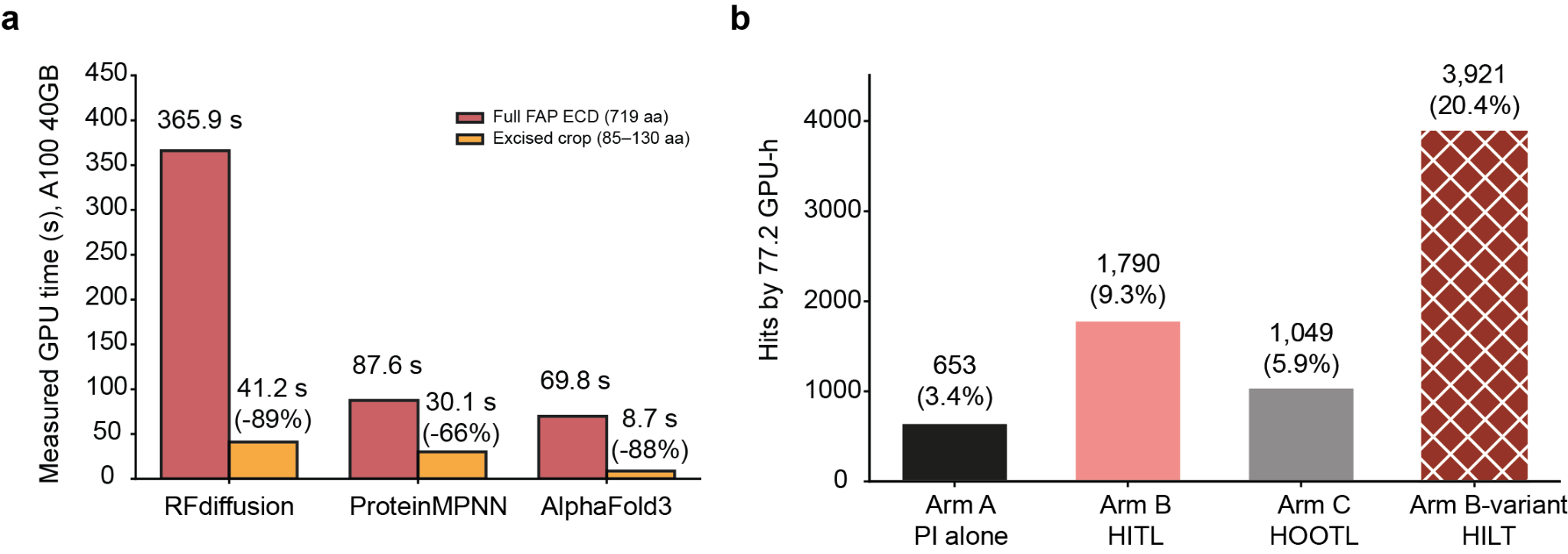}
\caption*{\textbf{Supporting Figure 5 (related to Figure 5)} | \textbf{(a)} Measured GPU time per stage, excised crop vs full FAP. Mean wall clock time on a single NVIDIA A100 40 GB for each pipeline stage, run against the full 719-residue FAP ectodomain or an 85-residue excised crop for one epitope cluster (hotspots 654/707/728). RFdiffusion is shown per binder. Protein MPNN is shown per run of 3 backbones, 8 sequences. AlphaFold3 is shown per fold. Percentages give the reduction in mean time from full FAP to the crop. Standard deviations were under 1\% of the mean and are not shown. \textbf{(b)} Cumulative hits for each arm when its campaign is cut off at $77.2$ GPU-h, the total used by the shortest arm (Arm B). Percentages give the hit rates. Within the time used, each arm generated 19,184 designs (A), 19,184 designs (B), 17,808 designs (C), and 19,195 designs (B-variant).}
\end{figure}

\end{document}